\documentclass{ws-procs9x6}

\begin{document}

\title{PAMELA AND ATIC ANOMALIES\\
IN DECAYING GRAVITINO DARK MATTER SCENARIO}

\author{$^{(a)}$KOJI ISHIWATA, $^{(b)}$SHIGEKI MATSUMOTO and
  $^{(a)}$TAKEO MOROI}

\address{$^{(a)}$Department of Physics, Tohoku University, 
Sendai 980-8578, Japan\\
$^{(b)}$Department of Physics, University of Toyama, 
Toyama 930-8555, Japan
}

\begin{abstract}
  Motivated by the recent results from the PAMELA and ATIC, we study
  the cosmic-ray electron and positron produced by the decay of
  gravitino dark matter.  We calculate the cosmic-ray electron and
  positron fluxes and discuss implications to the PAMELA and ATIC
  data.  In this paper, we will show that the observed anomalous
  fluxes by the PAMELA and ATIC can be explained in such a scenario.
  We will also discuss the synchrotron radiation flux from the
  Galactic center in such a scenario.
\end{abstract}

\keywords{Style file; \LaTeX; Proceedings; World Scientific Publishing.}

\bodymatter

\section{Introduction}\label{sec:intro}

In astrophysics, the existence of dark matter (DM) is almost
conclusive.  According to the recent survey of WMAP
\cite{Hinshaw:2008kr}, it accounts for 23 \% of the total energy
density in the universe.  In the standard model of particle physics,
however, there does not exist candidate for DM, which is one of the
reasons to call for beyond the standard model.  Supersymmetry (SUSY)
is a promising model which can give an answer to the question; in the
framework of SUSY, lightest superparticle (LSP) is a viable candidate
for DM.

The fluxes of high energy cosmic rays give information about the
properties of DM.  In the recent years, accuracy of the measurements
of the fluxes have been significantly improved.  In particular,
anomalous signals are reported by PAMELA \cite{Adriani:2008zr} and
ATIC \cite{:2008zzr} in the observations of cosmic-ray $e^{\pm}$.  The
PAMELA and ATIC results have attracted many attentions because the
anomalies may indicate an unconventional nature of DM.  In fact, there
have been a sizable number of DM models are proposed to explain the
anomalies after the announcements of the PAMELA and ATIC results.
Among them, especially in decaying DM scenarios, the observed
anomalies can be well explained with the appropriate choice of the
lifetime of DM especially in leptonically decaying scenarios. (For
works calculating cosmic-ray $e^{\pm}$, see \cite{Ishiwata:2008cv,
  Ishiwata:2009vx} and references therein.)

In usual supersymmetric scenario, $R$-parity conservation is assumed,
which protects LSP from decaying into standard model particles and
makes it a viable candidate of DM.  If we consider the case
that $R$-parity is violated, LSP is no loner stable; however, if
$R$-parity violation (RPV) is weak enough, the lifetime of the LSP can
be much longer than the present age of the universe and LSP can play
the role of DM\cite{Takayama:2000uz}.  In addition, when the
order of the RPV is properly chosen to give the lifetime of
$O(10^{26}\ {\rm sec})$, produced cosmic-ray positron gives excellent
agreement with PAMELA data \cite{Ishiwata:2008cv}.

On the other hand, synchrotron radiation from the decay of DM may give
constrains directly to scenarios explaining the PAMELA and ATIC
anomalies. Since DM decays into energetic $e^{\pm}$ under the magnetic
fields in our galaxy, synchrotron radiation is inevitably
induced. Importantly, the WMAP collaboration has observed the
radiation in the whole sky, so that the observation gives constraints
on the scenarios of the $e^{\pm}$ production due to the decay of DM in
the Galactic halo.

In this paper, we consider gravitino (donated as $\psi_{\mu}$) LSP in
RPV.  In the scenario, we calculate cosmic-ray $e^{\pm}$ and
synchrotron radiation flux induced by them, paying particular
attentions to PAMELA and ATIC anomalies.
We will see that the PAMELA and ATIC anomalies are simultaneously
explained if the lifetime of the gravitino DM is $O(10^{26}\
{\rm sec})$ and the mass is $\sim 1 - 2\ {\rm TeV}$
\cite{Ishiwata:2008cv,Ishiwata:2009vx}.  In addition, synchrotron
radiation from the Galactic center is comparable with or smaller than
the observation \cite{Ishiwata:2008qy}.



\section{The Scenario and Model Framework}\label{sec:scenario}
In this section, we briefly explain the cosmological aspects and the
model framework of $\psi_{\mu}$-DM scenario in RPV.
With RPV, the $\psi_{\mu}$ LSP becomes unstable and energetic positron
can be produced by the decay.  Even if the $\psi_{\mu}$ is unstable,
it can be DM if the RPV is weak enough so that the lifetime of the
gravitino $\tau_{3/2}$ is much longer than the present age of the
universe \cite{Takayama:2000uz,Buchmuller:2007ui}.  In fact, such a
scenario has several advantages.  In the $\psi_{\mu}$-LSP scenario
with RPV, the thermal leptogenesis \cite{Fukugita:1986hr} becomes
possible without conflicting the big-bang nucleosynthesis constraints.
In addition, the fluxes of the positron and $\gamma$-ray can be as
large as the observed values, and the anomalies in those fluxes
observed by the HEAT \cite{Barwick:1997ig} and the EGRET
\cite{Sreekumar:1997un} experiments, respectively, can be
simultaneously explained in such a scenario if $\tau_{3/2}\sim
O(10^{26}\ {\rm sec})$ \cite{Ishiwata:2008cu,Ibarra:2008qg}.

Here, let us consider the bi-linear RPV interactions.  Using the bases
where the mixing terms between the up-type Higgs and the lepton
doublets are eliminated from the superpotential, the relevant RPV
interactions are given by
\begin{eqnarray}
 {\cal L}_{\rm RPV} 
 = B_i \tilde{L}_i H_u + m^2_{\tilde{L}_i H_d} \tilde{L}_i H^*_d 
 + {\rm h.c.},
 \label{L_RPV}
\end{eqnarray}
where $\tilde{L}_i$ is left-handed slepton doublet in $i$-th
generation, while $H_u$ and $H_d$ are up- and down-type Higgs boson
doublets, respectively.  Then, the $\psi_{\mu}$ decays as
$\psi_\mu\rightarrow l_i^\pm W^\mp$, $\nu_i Z$, $\nu_i h$, and
$\nu_i\gamma$, where $l_i^\pm$ and $\nu_i$ are the charged lepton and
the neutrino in $i$-th generation, respectively.  Taking account of
all the relevant Feynman diagrams, we calculate the branching ratios
of these processes \cite{Ishiwata:2008cu}.  When the gravitino mass
$m_{3/2}$ is larger than $m_W$, the dominant decay mode is
$\psi_\mu\rightarrow l_i^\pm W^\mp$.  In such a case, we see
$\tau_{3/2}\simeq 6\times 10^{25}\ {\rm sec}\times
(\kappa_i/10^{-10})^{-2} (m_{3/2}/1\ {\rm TeV})^{-3}$, where
$\kappa_i=(B_i\sin\beta + m^2_{\tilde{L}_i
  H_d}\cos\beta)/m_{\tilde{\nu}_i}^2$ is the ratio of the vacuum
expectation value of the sneutrino field to that of the Higgs boson,
with $\tan\beta =\langle H^0_u \rangle / \langle H^0_d \rangle$, and
$m_{\tilde{\nu}_{i}}$ being the sneutrino mass.  Thus, $\tau_{3/2}$ is
a free parameter and can be much longer than the present age of the
universe if the RPV parameters $B_i$ and $m^2_{\tilde{L}_i H_d}$ are
small enough.

\section{ Electron and Positron Fluxes}\label{sec:flux}
Let us first summarize our procedure to calculate the $e^{\pm}$ fluxes
$\Phi_{e^{\pm}}$.  (For detail, see
\cite{Ishiwata:2008cv,Ishiwata:2009vx,Ishiwata:2008cu}.)  We solve the
diffusion equation to take account of the effects of the propagation
of $e^{\pm}$.  The energy spectrum of the $e^{\pm}$ from DM
$f_{e^{\pm}}(E,\vec{r})$ evolves as \cite{Baltz:1998xv}
\begin{eqnarray}
  \frac{\partial  f_{e^{\pm}}}{\partial t}
  = K(E) \nabla^2 f_{e^{\pm}}
  + \frac{\partial}{\partial E}\left[ b(E) f_{e^{\pm}} \right]
  + Q.
 \label{diffeq}
\end{eqnarray}
The function $K$ is expressed as $K=K_0 E_{\rm GeV}^\delta$
\cite{Delahaye:2007fr}, where $E_{\rm GeV}$ is the energy in units of
GeV, while $b=1.0\times 10^{-16}\times E_{\rm GeV}^2\ {\rm GeV/sec}$.
In our numerical calculation, we use the following three sets of the
model parameters, called MED, M1, and M2 models, which are defined as
$(\delta, K_0[{\rm kpc^2/Myr}], L[{\rm kpc}])=(0.70,0.0112,4)$ (MED),
$(0.46,0.0765,15)$ (M1), and $(0.55,0.00595,1)$ (M2), with $R=20\ {\rm
  kpc}$ for all models.  Here, $L$ and $R$ are the half-height and the
radius of the diffusion zone, respectively.  The MED model is the
best-fit to the boron-to-carbon ratio analysis, while the maximal and
minimal positron fractions for $E\gtrsim 10\ {\rm GeV}$ are expected
to be estimated with M1 and M2 models, respectively.  We found that
the MED and M1 models give similar positron fraction, so only the
results with the MED and M2 models are shown in the following.
The source term is given as,
\begin{eqnarray}
 Q_{\rm dec} = \frac{1}{\tau_{\rm DM}} 
 \frac{\rho_{\rm DM}(\vec{x})}{m_{\rm DM}}
 \left[ \frac{dN_{e^{\pm}}}{dE} \right]_{\rm dec},
\end{eqnarray}
where $\tau_{\rm DM}$ is the lifetime of DM.  In the above
expressions, $[dN_{e^+}/dE]_{\rm dec}$ is the energy distributions of
the $e^{\pm}$ from single decay processes, respectively, and are
calculated by using PYTHIA package \cite{Sjostrand:2006za} for each DM
candidate.  In addition, $\rho_{\rm DM}$ is the DM mass density for
which we adopt the Navarro-Frank-White (NFW) mass density profile
\cite{Navarro:1996gj}: $\rho_{\rm NFW} (\vec{x}) = \rho_\odot r_\odot
(r_c + r_\odot)^2/r (r_c + r)^2$, where $\rho_\odot\simeq 0.30\ {\rm
  GeV/cm^3}$ is the local halo density around the solar system, $r_c
\simeq 20\ {\rm kpc}$ is the core radius of the DM profile, $r_\odot
\simeq 8.5\ {\rm kpc}$ is the distance between the Galactic center and
the solar system, and $r$ is the distance from the Galactic center.

Once $f_{e^\pm}$ are given by solving the above equation, the fluxes
can be obtained as $\left[ \Phi_{e^\pm} (E) \right]_{\rm DM} =
\frac{c}{4 \pi} f_{e^\pm}(E, \vec{x}_{\odot})$, where
$\vec{x}_{\odot}$ is the location of the solar system, and $c$ is the
speed of light. In order to calculate the total fluxes of $e^\pm$, we
also have to estimate the background fluxes. In our study, we adopt
the following fluxes for cosmic-ray $e^{\pm}$ produced
by collisions between primary protons and interstellar medium in our
galaxy \cite{Moskalenko:1997gh,Baltz:1998xv}:$[\Phi_{e^-}]_{\rm BG}=0.16 E_{\rm
  GeV}^{-1.1}/(1+11E_{\rm GeV}^{0.9}+3.2E_{\rm GeV}^{2.15}) +
0.70E_{\rm GeV}^{0.7}/(1+110E_{\rm GeV}^{1.5}+600E_{\rm
  GeV}^{2.9}+580E_{\rm GeV}^{4.2})\ {\rm GeV}^{-1}\ {\rm cm}^{-2} \
{\rm sec}^{-1}\ {\rm str}^{-1}$ for the electron, and
$[\Phi_{e^+}]_{\rm BG}=4.5E_{\rm GeV}^{0.7}/(1+650E_{\rm
  GeV}^{2.3}+1500E_{\rm GeV}^{4.2})\ {\rm GeV}^{-1}\ {\rm cm}^{-2} \
{\rm sec}^{-1}\ {\rm str}^{-1}$ for the positron.

\section{Synchrotron Radiation: formalism and the
  observation}\label{sec:synch}
In this section, we first show the formalism for calculation of
synchrotron radiation flux. (The detail is in \cite{Ishiwata:2008qy}.)
Then, we address the implication of the present observation of
synchrotron radiation from the Galactic center region.

Synchrotron radiation energy density per unit time and unit frequency
is expressed as
\begin{eqnarray}
  L_{\nu}(\vec{x})
  = \int dE \  {\cal P}(\nu,E) f_{e^{\pm}}(E,\vec{x}).
\end{eqnarray}
Here, ${\cal P}(\nu,E)$ is synchrotron radiation energy per unit time
and unit frequency from single $e^{\pm}$ with energy $E$.
Adopting the Galactic magnetic flux density of $B\sim3 \ \mu$G, we can
see that the synchrotron radiation in the observed frequency band of
the WMAP (i.e, $22-93$ GHz) is from the $e^{\pm}$ with the energy of $E
\sim 10-100$ GeV.
For the $e^{\pm}$ in such an energy range, $f_{e^{\pm}}$ can be
well approximated by
\begin{eqnarray}
  f_{e^{\pm}}^{\rm (local)} (E,\vec{x})
  = \frac{1}{\tau_{\rm DM}}
  \frac{\rho_{\rm DM}(\vec{x})}{m_{\rm DM}}
  \frac{ Y_{e^{\pm}}(>E)}{b(E,{\vec{x}})},
  \label{eq:f_e}
\end{eqnarray}
where $Y_{e^{\pm}} (>E) \equiv \int_E^{\infty} dE^{\prime}[
dN_{e^{\pm}}/dE^{\prime}]_{\rm dec}$.  Thus, we use
$f_{e^{\pm}}^{\rm (local)}$ in the calculation of synchrotron
radiation flux. Here, we note in this formula that we take into
account the effects of both synchrotron radiation and inverse Compton
scattering for energy loss rate as $b(E,\vec{x}) = P_{\rm
  synch}+P_{\rm IC}$.  This is because, in the Galactic center region,
the inverse Compton scattering in the infrared $\gamma$-ray from stars
becomes the dominant energy-loss process, thus it can not be neglected.

In order to calculate the observed radiation energy flux, we integrate
$L_{\nu}(\vec{x})$ along the line of sight (l.o.s.), whose direction is
parametrized by the parameters $\theta$ and $\phi$, where $\theta$ is
the angle between the direction to the Galactic center and that of the
line of sight, and $\phi$ is the rotating angle around the direction
to the Galactic center.  (The Galactic plane corresponds to $\phi=0$
and $\pi$.)  Then, the synchrotron radiation flux is given by
\begin{eqnarray}
  J_{\nu} (\theta, \phi)
  = \frac{1}{4 \pi} \int_{\rm l.o.s.} d \vec{l} L_{\nu}(\vec{l}).
\end{eqnarray}
Notice that, adopting the approximation of the constant magnetic flux
in the Galaxy, the line of sight and energy integrals factorize.

Radiation flux from Galactic center region has been observed by the
WMAP for frequency bands of 22, 33, 41, 61, and 93 GHz
\cite{Dobler:2007wv, Gold:2008kp}.  Since then, intensive analysis has
been performed to understand the origins of the radiation flux.  (For
recent studies, see \cite{Dobler:2007wv, Hooper:2007kb, Gold:2008kp}.)
Most of the radiation flux is expected to be from astrophysical
origins, such as thermal dust, spinning dust, ionized gas, and
synchrotron radiation, which have been studied by the use of other
survey data \cite{Finkbeiner:2003im}.  With the three-year data, the
WMAP collaboration claimed that the flux intensity can be explained by
the known astrophysical origins \cite{Hinshaw:2006ia}.  On the
contrary, Refs.\ \cite{Hooper:2007kb, Dobler:2007wv} also studied the
WMAP three-year data, and claimed that there exists a remnant flux
from unknown origin which might be non-astrophysical; the remnant flux
is called the ``WMAP Haze''.
However, no clear indication of the WMAP Haze from unknown source was
reported by the WMAP collaboration after five-year data
\cite{Gold:2008kp}.

The existence of the WMAP Haze seems still controversial, and the
detailed studies of the WMAP Haze using the data is beyond the scope
of our study.  Here, we adopt the flux of the WMAP Haze suggested in
\cite{Hooper:2007kb} (i.e. $O(1~{\rm kJy/str})$) as a reference value.

\section{Numerical results}\label{sec:results}

\begin{figure}[t]
   \begin{center}
     \epsfxsize=0.6\textwidth\epsfbox{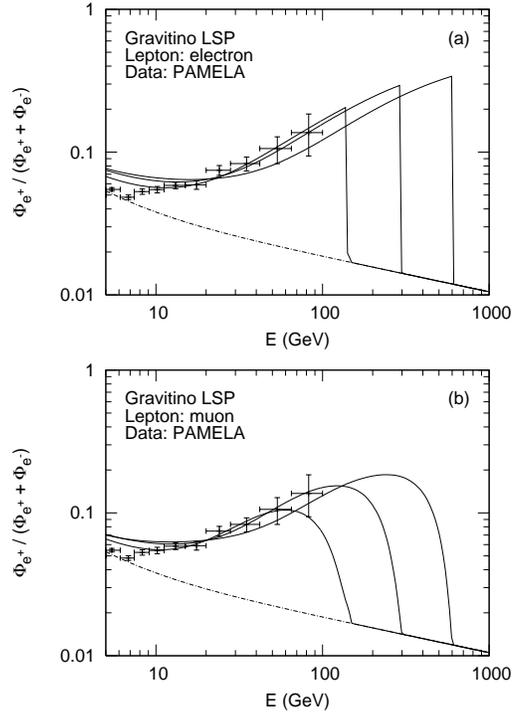}
     \caption{Positron fractions for the case that $\psi_{\mu}$
       dominantly decays to (a) the first-generation lepton in MED
       model and (b) the second-generation lepton in M2 model. Here,
       we take $m_{3/2}=300\ {\rm GeV}$, $600\ {\rm GeV}$, and $1.2\
       {\rm TeV}$ (from left to right), with $\tau_{3/2}=2.0\ \times
       10^{26}\ {\rm sec}$, $1.1\ \times 10^{26}\ {\rm sec}$, and
       $8.6\ \times 10^{25}\ {\rm sec}$ ($9.3\ \times 10^{25}\ {\rm
         sec}$, $5.8\ \times 10^{25}\ {\rm sec}$, and $5.0\ \times
       10^{25}\ {\rm sec}$) in (a) ((b)), respectively. Dot-dashed
       line is the fraction calculated only by the background fluxes.
     }
     \label{fig:grav_elmu}
   \end{center}
   \vspace{-0.5cm}
\end{figure}


First, we show the numerical results of the positron fraction.  For
simplicity, assuming a hierarchy among the RPV coupling constants, we
consider the case where the $\psi_{\mu}$ decays selectively into the
lepton in one of three generations (plus $W^\pm$, $Z$, or $h$).  In
Fig. \ref{fig:grav_elmu}, we show the positron fraction for the case
that the $\psi_{\mu}$ decays only into first- (second-) generation
lepton.  Here, we use MED (M2) model for first- (second-) generation
case and take $m_{3/2}=300$ GeV, 600 GeV, and 1.2 TeV, with
$\tau_{3/2}=2.0\ \times 10^{26}\ {\rm sec}$, $1.1\ \times 10^{26}\
{\rm sec}$, and $8.6\ \times 10^{25}\ {\rm sec}$ ($9.3\ \times
10^{25}\ {\rm sec}$, $5.8\ \times 10^{25}\ {\rm sec}$, and $5.0\
\times 10^{25}\ {\rm sec}$), which are the best-fit lifetime with
PAMELA data, respectively.  Here, in order to determine the best-fit
lifetime, we calculate $\chi^2$ by the use of PAMELA data.  (In our
$\chi^2$ analysis, since the positron fraction in the low energy
region is sensitive to the background fluxes, we only use the data
points with $E \geq 15$ GeV.) From the figure, we see that the
positron fraction well agrees with the PAMELA data for $m_{3/2}\gtrsim
100\ {\rm GeV}$ irrespective of the gravitino mass if $\tau_{3/2}$ is
properly chosen.  (Simultaneously, the energetic $\gamma$-ray flux is
also enhanced, which can be an explanation of the $\gamma$-ray excess
observed by the EGRET \cite{Sreekumar:1997un}.)

\begin{figure}[t]
  \begin{center}
    \epsfxsize=0.6\textwidth\epsfbox{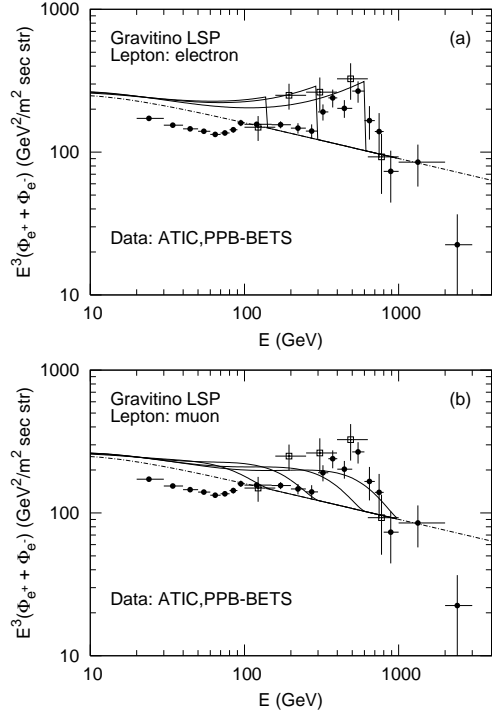}
    \caption{Total flux: $\Phi_{e^+}+\Phi_{e^-}$ with MED (M2) model
      for the case that the $\psi_{\mu}$ dominantly decays to the
      first- (second-) generation lepton in (a) ((b)). Dot-dashed line
      is the background flux.  Here, we use the same parameters in (a)
      and (b) of Fig. \ref{fig:grav_elmu}, respectively. In (b), we
      also plot the flux with $m_{3/2}=2$ TeV and $\tau_{3/2}=4.6
      \times 10^{25}$ sec and PPB-BETS data \cite{Torii:2008xu} . }
    \label{fig:totalflux}
  \end{center}
\end{figure}

Next, we move on to the total flux: $\Phi_{e^+}+\Phi_{e^-}$.  The
numerical results are shown in Fig.\ref{fig:totalflux}. Here, we use
the best-fit lifetime with the PAMELA data for each $m_{3/2}$, namely
the same value in (a) and (b) of Fig. \ref{fig:grav_elmu},
respectively.  From the figure, we see that the observed anomalous
structure is well reproduced in the both cases.  Especially, the
result is a good agreement with the observation when $m_{3/2}=1.2$ TeV
(2 TeV) for the case that the final state lepton is the first-
(second-) generation.  We also note that, in the total flux, the
numerical results does not change drastically by the choice of the
background.  This is because the signal from the gravitino is larger
than (or at least comparable to) the background.

\begin{figure}[t]
  \begin{center}
    \epsfxsize=0.9\textwidth\epsfbox{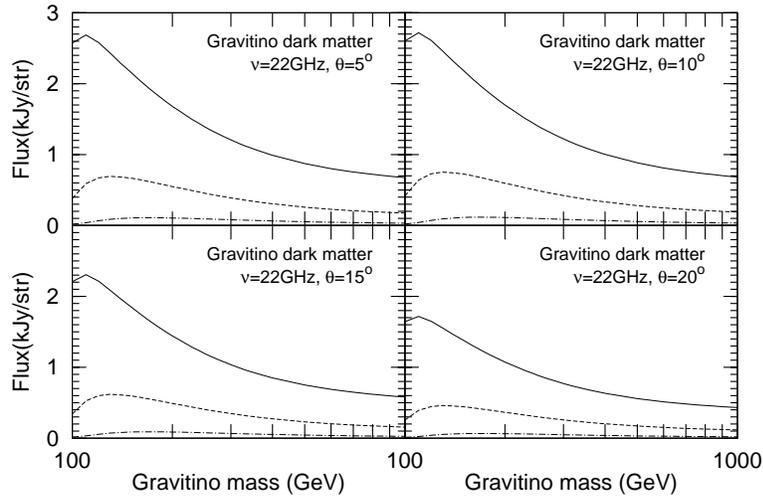}
    \caption{Synchrotron radiation fluxes at $\nu=22\ {\rm GHz}$ as
      functions of gravitino mass for angle $\theta = 5^{\circ}$,
      $10^{\circ}$, $15^{\circ}$, and $20^{\circ}$.  The final-state
      lepton in the $\psi_{\mu}$ decay is in the first generation.  Here,
      we take $\tau_{3/2}=5 \times 10^{26}$ sec, and show the cases of
      $B=$1, 3, 10 $\mu$G (from the bottom to the top) for each
      figure.}
    \label{fig:fluxgr2e_22}
  \end{center}
\end{figure}

Finally, let us discuss the synchrotron radiation flux.  The numerical
results are shown in Fig.\ref{fig:fluxgr2e_22}.  In this figure, we
consider the case that the $\psi_{\mu}$ mainly decays to first generation
lepton and plot for $\nu=22$ GHz as the function of $m_{3/2}$, taking
$\tau_{3/2}=5 \times 10^{26}$ sec.  The angle is set as
$\phi=\frac{\pi}{2}$, and $\theta=5^{\circ},~10^{\circ},~15^{\circ},$
and $20^{\circ}$, and we take $B=1,~3,$ and 10 $\mu$G.  For the
$\psi_{\mu}$-DM case, it can be seen that the synchrotron
radiation flux is of the order of $\sim 1\ {\rm kJy/str}$ or smaller.
As we mentioned, since the the existence of the exotic radiation flux
of this size is controversial, it is difficult to confirm or exclude
the present scenario without better understandings of the sources of
Galactic foreground emission.

\section{Conclusions}
In this paper, we have studied the cosmic-ray fluxes from the
$\psi_{\mu}$-DM decay in RPV, motivated by the recent observations by
PAMELA and ATIC.  Assuming that the $\psi_{\mu}$ is the dominant
component of DM, we calculate the cosmic-ray $e^{\pm}$, and found that
the both anomalies can be well explained when $\tau_{3/2} \sim
O(10^{26}~{\rm sec})$.  In particular, we saw that the ATIC anomaly
indicates $m_{3/2} \sim 1-2$ TeV in this scenario.  We also calculate
the synchrotron radiation induced by the cosmic-ray $e^{\pm}$ from
the Galactic center region with the lifetime to explain PAMELA and
ATIC anomalies.  Then, we obtained the result that the synchrotron
radiation flux is $O(1\ {\rm kJy/str})$ or smaller, which does not
exclude our scenario by the observation of the Galactic foreground
emission.

\bibliographystyle{ws-procs9x6}
\bibliography{ws-pro-sample}

\end{document}